\documentclass{article}
\usepackage{spconf,amsmath,graphicx,hyperref}
\usepackage{cite}
\usepackage{algorithmic}
\usepackage{amssymb}
\usepackage{graphicx}
\usepackage{textcomp}
\usepackage{xcolor}
\usepackage{booktabs}
\usepackage{makecell}
\usepackage{color}
\usepackage{colortbl}
\usepackage{multirow}
\usepackage{multicol}
\usepackage{adjustbox}
\usepackage{arydshln}
\usepackage[flushleft]{threeparttable}
\definecolor{SkyBlueRow}{RGB}{240,248,255}
\definecolor{myblue}{RGB}{108, 142, 191}
\hypersetup{
    colorlinks=true,
    linkcolor=myblue,
    citecolor=myblue,
    filecolor=myblue,
    urlcolor=myblue
}

\title{Boundary and Intra-Segment Learning for Partial Audio Deepfake Localization}
\name{Zhe Ye$^{1,4}$, Xiangui Kang$^{1,\dagger}$, Minhua Huang$^2$, Kai Wu$^2$, Kong Aik Lee$^3$, Chng Eng Siong$^4$\thanks{$\dagger$ Corresponding author.}}
\address{$^1$ Guangdong Key Lab of Information Security, \\ School of Computer Science and Engineering, Sun Yat-sen University, China, \\
$^2$ China Mobile Internet Corporation, China,
$^3$ The Hong Kong Polytechnic University, Hong Kong,
\\
$^4$ Nanyang Technological University, Singapore}
\begin{document}
%
\maketitle
\begin{abstract}
Partial audio deepfakes manipulate only selected speech regions, making them difficult to be localized. Existing methods exploit boundary cues for partial deepfake localization, but primarily focus on identifying boundary positions rather than modeling the feature changes that characterize authenticity transitions. Meanwhile, the internal characteristics of continuous bona fide and spoofed segments remain underexplored. In this paper, we propose Boundary and Intra-Segment Learning (BISL), which introduces boundary learning to model feature differences between adjacent frames and distinguish authenticity transitions from general acoustic variations. In addition, intra-segment learning captures the overall characteristics of continuous bona fide and spoofed segments while enhancing feature consistency within each segment. By jointly learning frame, boundary, and segment information, BISL enables more effective fine-grained partial audio deepfake localization. Experiments on multiple localization benchmarks show that BISL achieves an EER of 2.52\% and an F1-score of 97.40\% on PartialSpoof, outperforming the compared methods, while maintaining competitive performance on HAD and improved cross-dataset performance on LPS. The code will be made publicly available upon acceptance.
\end{abstract}
\section{Introduction}

The rapid development of generative AI has raised growing security concerns, including identity impersonation, misinformation, and threats on voice authentication systems \cite{ye2025speed, bakari2026identity, 11460864, 11296926 }. In particular, modern Text-to-Speech (TTS) and Voice Conversion (VC) systems \cite{guo2026qe,yin2026dmp} can generate highly natural speech with strong speaker similarity, further increasing the risk of audio deepfakes. To address these threats, numerous studies have developed effective countermeasures for audio deepfake detection \cite{ye2026breathnet, truong2026addressing}. However, attackers can manipulate only selected regions of an utterance while preserving the remaining bona fide content, resulting in partially spoofed audio. Such manipulation poses greater challenges for localization, as bona fide and spoofed content coexist within the same utterance and must be distinguished at the frame level.

Existing partial audio deepfake localization methods commonly perform frame-level detection to predict the authenticity of different temporal frames\cite{zhang2021initial, zhang2022partialspoof, yadav2024mdrt, cai2024integrating, wu2024coarse, ge2025gncl, zeng2025adversarial}. To further improve localization, Temporal Deepfake Location (TDL) \cite{xie2024efficient} focuses on temporal relationships between neighboring frames, using embedding similarity learning to improve the discrimination between bona fide and spoofed frames and temporal convolution to aggregate useful information from neighboring frames. Boundary-aware Attention Mechanism (BAM) \cite{zhong2024enhancing} extracts boundary features by combining intra-frame and inter-frame information and uses boundary predictions to control feature interactions between frames. Boundary-Frame Cross Graph Attention Network (BFC-Net) \cite{zhou2025bfc} further combines boundary features with inter-frame differences and uses boundary predictions to guide inter-frame interactions. Segment-Aware Learning (SAL) \cite{mao2026localizing} focuses on continuous speech segments by introducing segment positional labels to describe the position of each frame within a segment and cross-segment mixing to reduce over-reliance on boundary artifacts. However, the overall characteristics of continuous bona fide and spoofed regions remain underexplored, while the feature changes between adjacent frames during authenticity transitions have not been adequately captured.

\begin{figure*}[t]
\centering
\includegraphics[width=\textwidth]{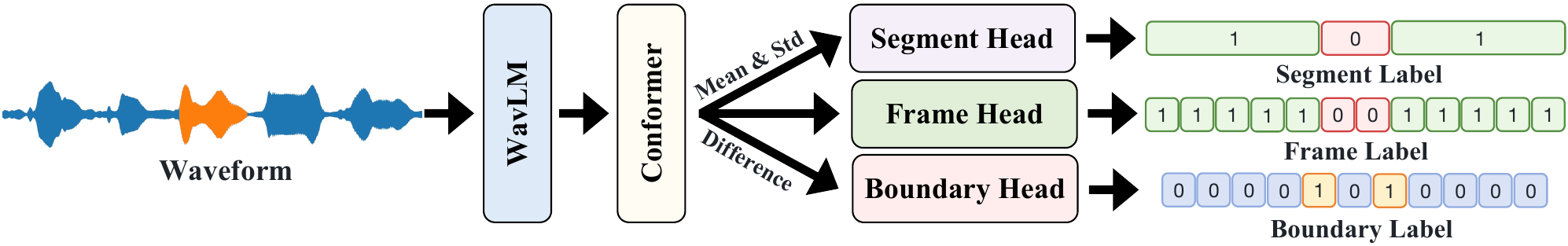}
\caption{Overview of BISL. Frame features extracted by WavLM and a Conformer are jointly supervised by frame-level classification, segment-level classification using temporal mean and standard deviation, and boundary classification using adjacent-frame feature differences. Segment and boundary heads are used only during training. All three heads share the same two-layer MLP architecture.}
\label{fig-pipeline}
\end{figure*}

Based on these observations, we propose Boundary and Intra-Segment Learning (BISL) for partial audio deepfake localization, which incorporates boundary learning and intra-segment learning. Specifically, boundary learning models the feature differences between adjacent frames, with boundary labels derived from changes in their authenticity labels. This encourages the model to distinguish feature changes associated with authenticity transitions from general acoustic variations. For intra-segment learning, we construct continuous bona fide and spoofed segments based on frame-level labels and aggregate the frame features within each segment for segment-level authenticity classification, enabling the model to capture the overall characteristics of each continuous region beyond frame-level prediction. We further employ an intra-segment compactness loss to encourage frame features within each segment to move toward the segment center, enhancing feature consistency and providing more stable representations for segment-level classification. By jointly learning frame, boundary, and segment information, BISL integrates complementary information at different temporal levels for fine-grained partial audio deepfake localization.

Experimental results on multiple partial audio deepfake localization datasets demonstrate the effectiveness of BISL. On PartialSpoof (PS), BISL achieves state-of-the-art performance with an EER of 2.52\% and an F1-score of 97.40\%, while on Half-truth Audio Deepfake (HAD), it achieves competitive performance with an EER of 0.07\% and an F1-score of 99.97\%. Under cross-domain evaluation on LlamaPartialSpoof (LPS), BISL obtains an EER of 37.10\% and an F1-score of 53.61\%. Furthermore, ablation studies evaluate the contribution of each component.

\section{Methods}
Given an input waveform, we employ WavLM-Large\footnote{\url{https://github.com/microsoft/unilm/tree/master/wavlm}} as the acoustic encoder and aggregate its layer-wise representations through a learnable weighted sum. The fused features are subsequently processed by a two-layer Conformer for temporal modeling, resulting in frame-level features. Based on these features, we introduce BISL with boundary and intra-segment learning as auxiliary supervision for the primary frame-level detection task. Each task uses a two-layer MLP head with a 256-dimensional hidden layer for binary classification.

\subsection{Boundary Learning}

Boundary learning focuses on capturing authenticity transitions between adjacent frames, complementing frame-level classification that predicts the authenticity of individual frames. Given an utterance with $T$ frames, we denote the frame-level feature at the $t$-th frame as $\mathbf{h}_t$. We compute the feature difference between adjacent frames to characterize local changes in the learned features:

\begin{equation}
\mathbf{d}_t = \mathbf{h}_{t+1} - \mathbf{h}_t,
\quad t=1,\ldots,T-1.
\end{equation}

The boundary label indicates whether an authenticity transition occurs between adjacent frames and is derived directly from the corresponding frame-level labels as follows:

\begin{equation}
b_t =
\begin{cases}
1, & y_{t+1} \neq y_t, \\
0, & y_{t+1} = y_t,
\end{cases}
\quad t=1,\ldots,T-1.
\end{equation}
where $y_t$ denotes the label of the $t$-th frame, and $b_t$ is the corresponding boundary label. Accordingly, an utterance with $T$ frames yields $T-1$ boundary labels.

Each difference feature $\mathbf{d}_t$ is fed into the boundary classification head to obtain the corresponding boundary prediction $\mathbf{p}_t^{\mathrm{bd}}$. The boundary classification loss is defined as:
\begin{equation}
\mathcal{L}_{\mathrm{bd}}
=
\frac{1}{T-1}
\sum_{t=1}^{T-1}
\mathrm{CrossEntropy}
\left(
\mathbf{p}_{t}^{\mathrm{bd}}, b_t
\right).
\end{equation}

Boundary learning aims to capture feature changes that are specifically associated with authenticity transitions. However, changes between adjacent frames can also arise from variations in phonetic content or acoustic conditions and therefore do not necessarily indicate spoofing boundaries. By deriving boundary labels from frame-level labels, the model is explicitly guided to distinguish authenticity-related transitions from such general acoustic variations. This allows the model to better characterize spoofing boundaries and provides complementary information for frame-level detection.

\subsection{Intra-Segment Learning}

We introduce Intra-Segment Learning to enhance the model’s discrimination between bona fide and spoofed regions in continuous speech. Specifically, during training, we partition each utterance into continuous segments according to the ground-truth frame-level labels, such that all frames within each segment share the same authenticity label. To reduce the potential influence of adjacent segments near the boundaries, we exclude the first and last frames of each segment and use the remaining interior frames for intra-segment learning. For segments with insufficient interior frames, the complete segment is used directly for intra-segment learning.

For each continuous segment, we calculate the mean and standard deviation of the features of its interior frames along the temporal dimension and concatenate them to obtain the segment-level representation:

\begin{equation}
\mathbf{s}_i =
\left[
\operatorname{Mean}_{\tau}(\mathbf{h}_{i,\tau});
\operatorname{Std}_{\tau}(\mathbf{h}_{i,\tau})
\right],
\end{equation}
where $\mathbf{h}_{i,\tau}$ denotes the feature of the $\tau$-th frame in the $i$-th segment. 

The mean captures the overall characteristics of the segment, while the standard deviation characterizes its internal feature variation. Their combination integrates information across consecutive frames, providing a more stable representation of continuous bona fide and spoofed regions than independent frame-level features. The segment-level representation $\mathbf{s}_i$ is fed into the segment classification head to predict the authenticity label of each segment. The segment classification loss is defined as:

\begin{equation}
\mathcal{L}_{\mathrm{seg}}^{\mathrm{cls}}
=
\frac{1}{N_{\mathrm{cls}}}
\sum_{i=1}^{N_{\mathrm{cls}}}
\mathrm{CrossEntropy}
\left(
\mathbf{p}_{i}^{\mathrm{seg}}, y_{i}
\right),
\end{equation}
where $\mathbf{p}_{i}^{\mathrm{seg}}$ denotes the segment-level prediction
obtained from $\mathbf{s}_i$, $y_{i}$ is the corresponding
segment label, and $N_{\mathrm{cls}}$ is the number of valid segments.

Despite sharing the same authenticity label, frame features within a segment may exhibit substantial dispersion, which can reduce the consistency of the segment representation. To promote intra-segment compactness, for segments containing at least two frames, we first compute the mean of the frame features as the segment center:

\begin{equation}
\mathbf{c}_i =
\frac{1}{T_i}
\sum_{\tau=1}^{T_i}
\mathbf{h}_{i,\tau},
\end{equation}
where $T_i$ is the number of frames used for the $i$-th segment. 

We then minimize the cosine distance between each frame feature and its corresponding segment center:

\begin{equation}
\mathcal{L}_{\mathrm{seg}}^{\mathrm{com}}
=
\frac{1}{N_{\mathrm{com}}}
\sum_{i=1}^{N_{\mathrm{com}}}
\frac{1}{T_i}
\sum_{\tau=1}^{T_i}
\left(
1-
\frac{
\mathbf{h}_{i,\tau}^{\top}\mathbf{c}_i
}{
\|\mathbf{h}_{i,\tau}\|_2
\|\mathbf{c}_i\|_2
}
\right).
\end{equation}
where $N_{\mathrm{com}}$ is the number of valid segments.

This loss encourages frame features within the same continuous segment to cluster around their segment center, thereby enhancing intra-segment feature consistency.

\subsection{Multi-task Training Objective}

The proposed framework is jointly optimized using a primary frame-level detection objective together with two auxiliary objectives for intra-segment learning and boundary detection. Frame, segment, and boundary predictions are obtained through three classification heads with the same architecture.

The frame-level detection loss is defined as:
\begin{equation}
\mathcal{L}_{\mathrm{frame}}
=
\frac{1}{T}
\sum_{t=1}^{T}
\mathrm{CrossEntropy}
\left(
\mathbf{p}_{t}^{\mathrm{frame}}, y_t
\right),
\end{equation}
where $\mathbf{p}_{t}^{\mathrm{frame}}$ denotes the prediction for the $t$-th frame, $y_t$ is the frame-level label, and $T$ is the number of frames.

The overall training objective is formulated as
\begin{equation}
\mathcal{L}
=
\mathcal{L}_{\mathrm{frame}}
+
\lambda_{\mathrm{cls}}\mathcal{L}_{\mathrm{seg}}^{\mathrm{cls}}
+
\lambda_{\mathrm{com}}\mathcal{L}_{\mathrm{seg}}^{\mathrm{com}}
+
\lambda_{\mathrm{bd}}\mathcal{L}_{\mathrm{bd}},
\end{equation}
where $\lambda_{\mathrm{cls}}$, $\lambda_{\mathrm{com}}$, and $\lambda_{\mathrm{bd}}$ denote the weighting coefficients for the corresponding loss terms. Specifically, $\mathcal{L}_{\mathrm{seg}}^{\mathrm{cls}}$ and $\mathcal{L}_{\mathrm{seg}}^{\mathrm{com}}$ jointly constitute the intra-segment learning, whereas $\mathcal{L}_{\mathrm{bd}}$ corresponds to the boundary learning.

\section{Experiments and Analysis}
\subsection{Experimental Settings}

\textbf{Datasets:} Our experiments are conducted on three partially spoofed audio datasets: PartialSpoof (PS) \cite{zhang2022partialspoof}, Half-truth Audio Detection (HAD) \cite{yi2021half}, and LlamaPartialSpoof (LPS) \cite{luong2025llamapartialspoof}. We conduct in-domain evaluation on PS and HAD using their official splits. For cross-domain evaluation, the model trained only on PS is directly evaluated on the crossfade version of LPS without adaptation.


\noindent\textbf{Baselines:} We compare our method with several localization methods, including LCNN-BLSTM \cite{zhang2021initial}, Multi-resolution \cite{zhang2022partialspoof}, MDRT \cite{yadav2024mdrt}, SPF \cite{cai2024integrating}, CFPRF \cite{wu2024coarse}, GNCL \cite{ge2025gncl}, AGO \cite{zeng2025adversarial}, TDL \cite{xie2024efficient}, BAM \cite{zhong2024enhancing}, BFC-Net \cite{zhou2025bfc}, and SAL \cite{mao2026localizing}. 

\noindent\textbf{Implementation Details:} All audio is sampled at 16 kHz. During training, each utterance is padded or truncated to 4 seconds, whereas the full-length utterance is used for evaluation. Following prior work, the temporal resolution is set to 160 ms for PS and 20 ms for HAD and LPS. RawBoost \cite{tak2022rawboost} is applied with a probability of 0.2 for data augmentation. The models are trained for up to 50 epochs using the Adam optimizer with a batch size of 10, an initial learning rate of $1 \times 10^{-5}$, and a weight decay of $1 \times 10^{-4}$. The learning rate is reduced by a factor of 0.1 every 10 epochs using a StepLR scheduler. The loss weights $\lambda_{\mathrm{cls}}$, $\lambda_{\mathrm{com}}$, and $\lambda_{\mathrm{bd}}$ are set to 0.2, 0.5, and 1.0, respectively. Early stopping with a patience of 5 epochs is applied based on the training loss, and the selected checkpoint is used for final evaluation. All experiments are conducted on a single NVIDIA GeForce RTX 3090 GPU.

\noindent\textbf{Evaluation Metric:} We evaluate the localization performance using two frame-level metrics: Equal Error Rate (EER) and F1-score. A lower EER and a higher F1-score indicate better localization performance.

\begin{table}[t]
\centering
\setlength{\tabcolsep}{1mm}
\caption{Metrics (\%) of different methods trained and evaluated on PS. $^{\ddagger}$ denotes results re-implemented by us.}
\vspace{2mm}
\label{tab:ps}
\begin{tabular}{l|l|cc}
\Xhline{1pt}
\textbf{System} & \textbf{Front-end} & \textbf{EER $\downarrow$} & \textbf{F1 $\uparrow$} \\
\hline
LCNN-BLSTM\cite{zhang2021initial}      & LFCC              & 16.21     & -      \\
Multi reso.\cite{zhang2022partialspoof}  & W2V2-Large        & 9.24      & -      \\
MDRT\cite{yadav2024mdrt}            & W2V2-Base+M2D     & 8.82      & -      \\
SPF\cite{cai2024integrating}             & WavLM             & -         & 92.96  \\
CFPRF\cite{wu2024coarse}           & W2V2-XLSR         & 7.41      & 93.89  \\
GNCL\cite{ge2025gncl}            & W2V2-Base         & 11.81     & -      \\
AGO\cite{zeng2025adversarial}             & W2V2-XLSR         & 6.79      & 94.36  \\
TDL\cite{xie2024efficient}             & W2V2-XLSR         & 7.04      & 88.96  \\
BAM\cite{zhong2024enhancing}             & WavLM             & 3.58      & 96.09  \\
BFC-Net\cite{zhou2025bfc}         & WavLM             & \underline{2.73}      & 96.69  \\
SAL$^{\ddagger}$\cite{mao2026localizing}        & WavLM             & 3.07         & \underline{97.06}      \\
\hline
\rowcolor{SkyBlueRow} Ours & WavLM             & \textbf{2.52}          & \textbf{97.40}      \\
\Xhline{1pt}
\end{tabular}
\end{table}

\begin{table}[t]
\centering
\setlength{\tabcolsep}{2.5mm}
\caption{Metrics (\%) of different methods trained and evaluated on HAD. $^{\ddagger}$ denotes results re-implemented by us.}
\vspace{2mm}
\label{tab:had}
\begin{tabular}{l|l|cc}
\Xhline{1pt}
\textbf{System} & \textbf{Front-end} & \textbf{EER $\downarrow$} & \textbf{F1 $\uparrow$} \\
\hline
Multi reso.\cite{zhang2022partialspoof} & W2V2-Large   & 0.18         & 99.89 \\
SPF\cite{cai2024integrating}         & WavLM        & 0.35         & 99.78 \\
CFPRF\cite{wu2024coarse}               & W2V2-XLSR    & 0.08         & 99.95 \\
SAL$^{\ddagger}$\cite{mao2026localizing} & WavLM & \textbf{0.06} & \textbf{99.97} \\
\hline
\rowcolor{SkyBlueRow} 
Ours  & WavLM  & \underline{0.07} & \textbf{99.97}  \\    
\Xhline{1pt}
\end{tabular}
\end{table}

\subsection{Main Results}

We compare our method with existing partial spoofing detection methods on the PS, HAD, and LPS datasets. PS and HAD are used for in-domain evaluation, while LPS is used to evaluate cross-dataset generalization. 

As shown in Tables~\ref{tab:ps} and~\ref{tab:had}, our method achieves strong in-domain performance on both PS and HAD. On the PS dataset, our method achieves the best performance with an EER of 2.52\% and an F1-score of 97.40\%. Compared with the previous state-of-the-art results, our method achieves a relative EER reduction of 7.69\% and a relative improvement of 0.34\% in F1-score. On the HAD dataset, where most existing methods already achieve near-saturated performance, our method obtains an EER of 0.07\% and an F1-score of 99.97\%, achieving performance comparable to the state-of-the-art results. 

As shown in Table~\ref{tab:lps}, under the cross-dataset setting, where models are trained on PS and directly evaluated on LPS, all methods exhibit a substantial performance degradation compared with their in-domain results, indicating a considerable distribution gap between PS and LPS. Despite this distribution shift, our method achieves the lowest EER of 37.10\% and the highest F1-score of 53.61\%. These results demonstrate the improved cross-dataset generalization of the proposed method to unseen data distributions.

Overall, our method achieves the best performance on PS and in the cross-dataset evaluation on LPS, while remaining competitive with the state-of-the-art methods on HAD.
\begin{table}[t]
\centering
\setlength{\tabcolsep}{6.5mm}
\caption{Metrics (\%) of different methods trained on PS and evaluated on LPS. $^{\ddagger}$ denotes results re-implemented by us.}
\vspace{2mm}
\label{tab:lps}
\begin{tabular}{l|cc}
\Xhline{1pt}
\textbf{System} & \textbf{EER $\downarrow$} & \textbf{F1 $\uparrow$} \\
\hline
Multi reso.\cite{zhang2022partialspoof} & 47.49 & -  \\
SAL$^{\ddagger}$\cite{mao2026localizing}   &  39.09   & 52.43     \\
\hline
\rowcolor{SkyBlueRow} 
Ours      &  \textbf{37.10}    & \textbf{53.61}      \\
\Xhline{1pt}
\end{tabular}
\end{table}

\begin{table}[t]
\centering
\setlength{\tabcolsep}{0.6mm}
\caption{Ablation study of boundary and intra-segment learning on the PS and LPS datasets.}
\vspace{2mm}
\label{tab:ablation}
\begin{tabular}{cc|cc|cc}
\Xhline{1pt}
\multirow{2}{*}{\textbf{Boundary}} &
\multirow{2}{*}{\textbf{Intra-segment}} &
\multicolumn{2}{c|}{\textbf{PS}} &
\multicolumn{2}{c}{\textbf{LPS}} \\
\cline{3-6}
& &
\textbf{EER $\downarrow$} & \textbf{F1 $\uparrow$} &
\textbf{EER $\downarrow$} & \textbf{F1 $\uparrow$} \\
\hline
$\times$   & $\times$   & 2.86 & 97.11 & 37.84 & 52.53 \\
$\times$   & \checkmark & 2.71 & 97.21 & 37.60 & 53.01 \\
\checkmark & $\times$   & 2.59 & 97.34 & 37.79 & 53.09 \\
\rowcolor{SkyBlueRow}
\checkmark & \checkmark &
\textbf{2.52} & \textbf{97.40} &
\textbf{37.10} & \textbf{53.61} \\
\Xhline{1pt}
\end{tabular}
\end{table}

\subsection{Ablation Results}

The ablation results are shown in Table~\ref{tab:ablation}. Compared with training using only frame-level supervision, both boundary learning and intra-segment learning improve localization performance when applied individually. Their joint incorporation further improves performance on both datasets, achieving the best overall performance with an EER of 2.52\% and an F1-score of 97.40\% on PS, and an EER of 37.10\% and an F1-score of 53.61\% on LPS.  This improvement suggests that intra-segment characteristics provide complementary information to boundary learning, benefiting localization.

\section{Conclusion}
In this paper, we propose Boundary and Intra-Segment Learning (BISL). BISL introduces segment-level learning to capture the overall characteristics of continuous bona fide and spoofed regions while improving their internal feature consistency. In addition, boundary learning uses the feature differences between adjacent frames to distinguish changes related to authenticity transitions from general acoustic variations. By jointly learning frame-level, segment-level, and boundary information, BISL enables more effective fine-grained partial deepfake localization. Experiments across multiple localization datasets demonstrate the effectiveness of BISL for partial deepfake localization. Ablation results further validate the effectiveness of each component in BISL.

\small
\bibliographystyle{IEEEbib}
\bibliography{refs}

@inproceedings{tak2022rawboost,
  title={Rawboost: A raw data boosting and augmentation method applied to automatic speaker verification anti-spoofing},
  author={Tak, Hemlata and Kamble, Madhu and Patino, Jose and Todisco, Massimiliano and Evans, Nicholas},
  booktitle={ICASSP 2022-2022 IEEE International Conference on Acoustics, Speech and Signal Processing (ICASSP)},
  pages={6382--6386},
  year={2022},
  organization={IEEE}
}

@inproceedings{luong2025llamapartialspoof,
  title={Llamapartialspoof: An llm-driven fake speech dataset simulating disinformation generation},
  author={Luong, Hieu-Thi and Li, Haoyang and Zhang, Lin and Lee, Kong Aik and Chng, Eng Siong},
  booktitle={ICASSP 2025-2025 IEEE International Conference on Acoustics, Speech and Signal Processing (ICASSP)},
  pages={1--5},
  year={2025},
  organization={IEEE}
}

@inproceedings{zhang2021initial,
  title={An Initial Investigation for Detecting Partially Spoofed Audio},
  author={Zhang, Lin and Wang, Xin and Cooper, Erica and Yamagishi, Junichi and Patino, Jose and Evans, Nicholas},
  booktitle={Proc. Interspeech 2021},
  pages={4264--4268},
  year={2021}
}

@inproceedings{ge2025gncl,
  title={Gncl: A graph neural network with consistency loss for segment-level spoofed speech detection},
  author={Ge, Zirui and Xu, Xinzhou and Guo, Haiyan and Yang, Zhen and Schuller, Bj{\"o}rn},
  booktitle={ICASSP 2025-2025 IEEE International Conference on Acoustics, Speech and Signal Processing (ICASSP)},
  pages={1--5},
  year={2025},
  organization={IEEE}
}

@inproceedings{yadav2024mdrt,
  title={Mdrt: Multi-domain synthetic speech localization},
  author={Yadav, Amit Kumar Singh and Bhagtani, Kratika and Baireddy, Sriram and Bestagini, Paolo and Tubaro, Stefano and Delp, Edward J},
  booktitle={ICASSP 2024-2024 IEEE International Conference on Acoustics, Speech and Signal Processing (ICASSP)},
  pages={11171--11175},
  year={2024},
  organization={IEEE}
}

@article{zhang2022partialspoof,
  title={The partialspoof database and countermeasures for the detection of short fake speech segments embedded in an utterance},
  author={Zhang, Lin and Wang, Xin and Cooper, Erica and Evans, Nicholas and Yamagishi, Junichi},
  journal={IEEE/ACM Transactions on Audio, Speech, and Language Processing},
  volume={31},
  pages={813--825},
  year={2022},
  publisher={IEEE}
}

@inproceedings{xie2024efficient,
  title={An efficient temporary deepfake location approach based embeddings for partially spoofed audio detection},
  author={Xie, Yuankun and Cheng, Haonan and Wang, Yutian and Ye, Long},
  booktitle={ICASSP 2024-2024 IEEE International Conference on Acoustics, Speech and Signal Processing (ICASSP)},
  pages={966--970},
  year={2024},
  organization={IEEE}
}

@article{cai2024integrating,
  title={Integrating frame-level boundary detection and deepfake detection for locating manipulated regions in partially spoofed audio forgery attacks},
  author={Cai, Zexin and Li, Ming},
  journal={Computer Speech \& Language},
  volume={85},
  pages={101597},
  year={2024},
  publisher={Elsevier}
}

@inproceedings{wu2024coarse,
  title={Coarse-to-fine proposal refinement framework for audio temporal forgery detection and localization},
  author={Wu, Junyan and Lu, Wei and Luo, Xiangyang and Yang, Rui and Wang, Qian and Cao, Xiaochun},
  booktitle={Proceedings of the 32nd ACM International Conference on Multimedia},
  pages={7395--7403},
  year={2024}
}

@inproceedings{zeng2025adversarial,
  title={Adversarial Training and Gradient Optimization for Partially Deepfake Audio Localization},
  author={Zeng, Siding and Yi, Jiangyan and Tao, Jianhua and He, Jiayi and Lian, Zheng and Liang, Shan and Zhang, Chuyuan and Chen, Yujie and Zhang, Xiaohui},
  booktitle={ICASSP 2025-2025 IEEE International Conference on Acoustics, Speech and Signal Processing (ICASSP)},
  pages={1--5},
  year={2025},
  organization={IEEE}
}

@inproceedings{zhong2024enhancing,
  title={Enhancing Partially Spoofed Audio Localization with Boundary-aware Attention Mechanism},
  author={Zhong, Jiafeng and Li, Bin and Yi, Jiangyan},
  booktitle={Proc. Interspeech 2024},
  pages={4838--4842},
  year={2024}
}

@article{zhou2025bfc,
  title={BFC-Net: Boundary-Frame cross graph attention network for partially spoofed audio localization},
  author={Zhou, Yi and Xue, Zhaodong and Senhadji, Lotfi and Shu, Huazhong and Wu, Jiasong},
  journal={Neurocomputing},
  pages={130867},
  year={2025},
  publisher={Elsevier}
}

@inproceedings{yi2021half,
  title={Half-Truth: A Partially Fake Audio Detection Dataset},
  author={Yi, Jiangyan and Bai, Ye and Tao, Jianhua and Ma, Haoxin and Tian, Zhengkun and Wang, Chenglong and Wang, Tao and Fu, Ruibo},
  booktitle={Proc. Interspeech 2021},
  pages={1654--1658},
  year={2021}
}

@inproceedings{mao2026localizing,
  title={Localizing Speech Deepfakes Beyond Transitions via Segment-Aware Learning},
  author={Mao, Yuchen and Huang, Wen and Qian, Yanmin},
  booktitle={ICASSP 2026 - 2026 IEEE International Conference on Acoustics, Speech and Signal Processing (ICASSP)},
  year={2026},
  organization={IEEE},
}

@ARTICLE{11296926,
  author={Ye, Zhe and Yan, Qiben and Chen, Jiahao and Kang, Xiangui and Huang, Jiwu},
  journal={IEEE Transactions on Information Forensics and Security}, 
  title={StealthPhase: Toward a Stealthy Backdoor Attack Against Speaker Recognition}, 
  year={2025},
  volume={20},
  number={},
  pages={13328-13341}
}

@INPROCEEDINGS{11460864,
  author={Dar, Daniyal Kabir and Yan, Qiben and Xiao, Li and Ross, Arun},
  booktitle={ICASSP 2026 - 2026 IEEE International Conference on Acoustics, Speech and Signal Processing (ICASSP)}, 
  title={Impact of Phonetics on Speaker Identity in Adversarial Voice Attack}, 
  year={2026},
  volume={},
  number={},
  pages={13462-13466},
}

@inproceedings{guo2026qe,
  title={QE-XVC: Zero-Shot Cross-Lingual Voice Conversion via Query-Enhancement and Conditional Flow Matching},
  author={Guo, Han-Jie and Du, Hui-Peng and Wang, Shi-Ming and Jiang, Xiao-Hang and Gao, Ying-Ying and Zhang, Shi-Lei and Ling, Zhen-Hua},
  booktitle={ICASSP 2026-2026 IEEE International Conference on Acoustics, Speech and Signal Processing (ICASSP)},
  pages={17342--17346},
  year={2026}
}

@inproceedings{yin2026dmp,
  title={DMP-TTS: Disentangled Multi-Modal Prompting for Controllable Text-to-Speech with Chained Guidance},
  author={Yin, Kang and Qiang, Chunyu and Zhao, Sirui and Wang, Xiaopeng and Liang, Yuzhe and Cai, Pengfei and Xu, Tong and Zhang, Chen and Chen, Enhong},
  booktitle={ICASSP 2026-2026 IEEE International Conference on Acoustics, Speech and Signal Processing (ICASSP)},
  pages={16477--16481},
  year={2026}
}

@article{ye2025speed,
  title={Speed master: Quick or slow play to attack speaker recognition},
  author={Ye, Zhe and Zhang, Wenjie and Ren, Ying and Kang, Xiangui and Yan, Diqun and Ma, Bin and Wang, Shiqi},
  journal={Proceedings of the AAAI Conference on Artificial Intelligence},
  volume={39},
  number={21},
  pages={22137--22145},
  year={2025}
}

@article{ye2026breathnet,
  title={BreathNet: Generalizable Audio Deepfake Detection via Breath-Cue-Guided Feature Refinement},
  author={Ye, Zhe and Kang, Xiangui and He, Jiayi and Chen, Chengxin and Zhu, Wei and Wu, Kai and Yang, Yin and Huang, Jiwu},
  journal={arXiv preprint arXiv:2602.13596},
  year={2026}
}

@inproceedings{truong2026addressing,
  title={Addressing Gradient Misalignment in Data-Augmented Training for Robust Speech Deepfake Detection},
  author={Truong, Duc-Tuan and Liu, Tianchi and Li, Junjie and Tao, Ruijie and Lee, Kong Aik and Chng, Eng Siong},
  booktitle={ICASSP 2026-2026 IEEE International Conference on Acoustics, Speech and Signal Processing (ICASSP)},
  pages={16252--16256},
  year={2026}
}

@inproceedings{bakari2026identity,
  title={Identity leakage through accent cues in voice anonymisation},
  author={Bakari, Rayane and Le Blouch, Olivier and Gengembre, Nicolas and Evans, Nicholas and Panariello, Michele},
  booktitle={ICASSP 2026-2026 IEEE International Conference on Acoustics, Speech and Signal Processing (ICASSP)},
  pages={13657--13661},
  year={2026}
}

\end{document}